\documentclass[journal,dvipsnames]{IEEEtran}
\ifCLASSINFOpdf
\else
\fi
\usepackage[utf8]{inputenc}
\usepackage{amsmath,amssymb,amsfonts}
\usepackage{cite}

\usepackage{tikz}
\usepackage[T1]{fontenc}
\usepackage{algorithmic}
\usepackage{stfloats}
\usepackage{graphicx}
\usepackage{subfigure}
\usepackage{textcomp}
\usepackage{xcolor}
\usepackage{pifont}
\usepackage{amsthm}

\usepackage{hyperref}
\usepackage{mathrsfs}
\usepackage{booktabs}
\usepackage{hyperref}

\usepackage{color}
\usepackage{cases}
\usepackage{comment}
\usepackage{empheq} 
\usepackage{caption}
\usepackage{subcaption}
\allowdisplaybreaks[4]
\usepackage{url}  
\usepackage{graphicx}  
\usepackage[ruled,vlined,linesnumbered]{algorithm2e}
\usepackage{setspace}
\usepackage{floatpag} 
\usepackage{float}
\floatpagestyle{empty}

\definecolor{red}{RGB}{0,0,0}

\usepackage{listings}

\begin{document}
\captionsetup[figure]{labelfont={rm},labelformat={default},labelsep=period,name={Fig.}}
\title{Movable Antenna Enabled ISAC Beamforming Design for Low-Altitude Airborne
Vehicles}
%
%
%

\author{Yue Xiu,~Songjie Yang,~Wanting Lyu,~Phee~Lep Yeoh,~\IEEEmembership{Senior Member,~IEEE},\\~Yonghui Li,~\IEEEmembership{Fellow,~IEEE},
~Yi Ai\\
\thanks{Yue Xiu and W.~Lyu are with 
National Key Laboratory of Science and Technology on Communications, University of Electronic Science and Technology of China, Chengdu 611731, China (E-mail:  
xiuyue12345678@163.com, lyuwanting@yeah.net).
Phee Lep Yeoh is with School of Science, Technology and Engineering, University of the Sunshine Coast, QLD, Australia(e-mail: pyeoh@usc.edu.au).
Yonghui Li is with the School of Electrical and Information Engineering, University of Sydney, Sydney, NSW 2006, Australia(e-mail: yonghui.li@sydney.edu.au).
Yi Ai is with College of Air Traffic Management, 
Civil Aviation Flight University of China, Deyang 618311, China.
}
\thanks{The corresponding author is Yi Ai.}}

\maketitle
\begin{abstract}
In mobile systems with low-altitude vehicles, integrated sensing and communication (ISAC) is considered an effective approach to increase the transmission rate due to limited spectrum resources. To further improve the ISAC performance, this paper proposes a novel method called integrated sensing and communication-movable antenna (ISAC-MA) to optimize the antenna's position. Our goal is to support low-space vehicles by optimizing radar and communication joint beamforming and antenna position in the presence of clutter. This scheme not only guarantees the required signal-to-noise ratio (SNR) for sensing but also further improves the SNR for communication. A successive convex approximation (SCA)-based block coordinate descent (BCD) algorithm is proposed to maximize communication capacity under the condition of sensing SNR. Numerical results show that, compared with the traditional ISAC system and various benchmark schemes, the proposed ISAC-MA system can achieve higher communication capacity under the same sensing SNR constraints.
\end{abstract}

\begin{IEEEkeywords}
Integrating sensing and communication, movable antenna, alternating optimization, successive convex approximation. 
\end{IEEEkeywords}

%
\IEEEpeerreviewmaketitle

\section{Introduction}
\IEEEPARstart{T}o provide high-speed and reliable services for users in remote areas in the sixth generation (6G) system, researchers have focused on deploying low-altitude networks\cite{wubben2021novel,8907440}. The space-air-ground integrated network (SAGIN) aims to balance the coverage, delay, and bandwidth requirements through the interoperation of space, air, and ground network segments. This integration is essential for addressing infrastructure dysfunction caused by natural disasters and malicious attacks. Mobility management and UAV system traffic management are crucial network functions for handling flight network components. They support air network deployment, ensure low-altitude aircraft safety, and manage UAV traffic. Generally, obtaining the location and motion information needed for mobility management relies on external infrastructure or auxiliary spaceborne inertial measurement units. Detecting mobile information while maintaining connectivity with internal resources is necessary to improve air networks' autonomy, which has driven the development of integrated sensing and communication (ISAC) in air networks.

ISAC is a new research topic in spectrum sharing \cite{10439221,9755276,9755276}. 
The common properties of signals used in both systems allow signal types suitable for one task to be used in another system, and common circuits in both systems reduce the cost of reshaping them. 
However, due to the limited beam resources, the performance of the ISAC system will degrade. In order to further improve the performance of the ISAC system and further explore the spatial degree of freedom (DOF) to improve the communication performance, a MA antenna system is proposed. Specifically, by connecting the MA to the radio frequency chain through a flexible cable, the position of the MA can be real-time adjusted by the controller\cite{10278220,10382559}. 
The ISAC-MA system can flexibly change the location of transmitting and receiving MA, so that the channel matrix between them can be reshaped to obtain higher capacity and sensing accuracy.

In this paper, we analyse the communication rate optimization of the ISAC-MA system with point target and clutter target fields that are low-altitude airborne vehicles. Specifically, we aim to optimize the transmit and receive MA positions and transmit beamforming to maximize system capacity under the radar sensing rate constraint. The simulation results show that MA can significantly improve the ISAC system's radar detection accuracy and communication rate. We futher highlight that the capacity improvement of ISAC-MA compared to ISAC without MA increases with increasing number of antennas. Furthermore, compared with existing optimization solutions based on gradient descent and zero-forcing, the proposed successive convex approximation (SCA)-based block coordinate descent (BCD) algorithm can achieve a higher channel capacity which increases with increasing transmit power.

\section{System Model and Problem Formulation}\label{II}
\begin{figure}[htbp]
  \centering
  \includegraphics[scale=0.6]{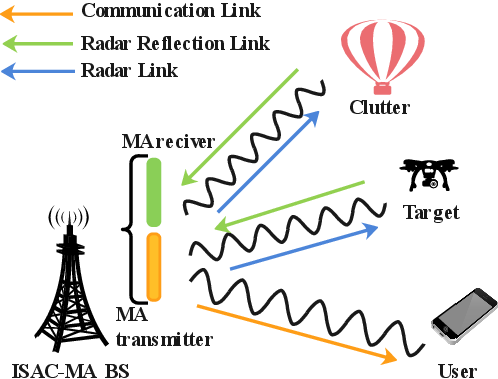}
  \caption{Illustation of the ISAC-MA System.}\vspace{-10pt}
\label{FIGURE0}
\end{figure}

In this letter, an ISAC communication system with $N_{T}$ transmit and $N_R$ receive MAs is considered, where the receiver is equipped with a single antenna, as shown in Fig.\ref{FIGURE0}. The index set of the transmit MAs is $\mathcal{N}=\{1,\ldots,N_{T}\}$. The coordinate of the $n$-th transmit MA is $x_{n}\in\mathcal{D}_{T}$, $n\in\mathcal{N}$, where $\mathcal{D}_{T}=[d_{T}|1\leq d_{T}\leq D_{T}]$ varies according to the moving range of the transmit MAs. Similarly, the $n$-th receive MA is $y_{n}\in\mathcal{D}_{R}$, $n\in\mathcal{N}$, where $\mathcal{D}_{R}=[d_{R}|1\leq d_{R}\leq D_{R}]$ is the transmit MAs moving range. In the ISAC system, the transmit data $s\in\mathbb{C}^{1\times 1}$ is transmitted by using transmit beamforming $\boldsymbol{w}\in\mathbb{C}^{N_{T}\times 1}$. Thus,
the reflection signal by point target and clutter target is expressed as
\begin{small}
\begin{align}
&\boldsymbol{r}=\sigma\boldsymbol{a}_{R}(\boldsymbol{y},\boldsymbol{\theta}^{P})\boldsymbol{a}_{T}^{H}(\boldsymbol{x},\boldsymbol{\theta}^{P})\boldsymbol{w}s+\delta\boldsymbol{a}_{R}(\boldsymbol{y},\boldsymbol{\theta}^{C})\boldsymbol{a}_{T}^{H}(\boldsymbol{x},\boldsymbol{\theta}^{C})\boldsymbol{w}s+\boldsymbol{n},\label{pro1}
\end{align}
\end{small}%
in which $\sigma$ and $\delta$ denote the target and clutter reflection factors. 
$\boldsymbol{a}_{\kappa_{1}}(\boldsymbol{\mu},\boldsymbol{\theta}^{\kappa_{2}})=\left[e^{j\frac{2\pi}{\lambda}\rho_{T}^{1}(\mu_{1},\theta_{1}^{\kappa_{2}})},e^{j\frac{2\pi}{\lambda}\rho_{T}^{2}(\mu_{2},\theta_{2}^{\kappa_{2}})}\right.$ $\left.,\ldots,e^{j\frac{2\pi}{\lambda}\rho_{T}^{N_{T}}(\mu_{N_{T}},\theta_{N_{T}}^{\kappa_{2}})}\right]\in\mathbb{C}^{N_{T}\times 1}$ is the transmit field response vector and receive field response vector, where $\kappa_{1}\in\{T,R\}$ is transmitter and receiver indication set, $\kappa_{2}\in\{P,C\}$ is target and clutter indication set, $\boldsymbol{\mu}\in\{\boldsymbol{x},\boldsymbol{y}\}$ is communication MA position and sensing MA position indication set, $\boldsymbol{x}=[x_{1},\ldots,x_{N_{T}}]^{T}$,$\boldsymbol{y}=[y_{1},\ldots,y_{N_{T}}]^{T}$, $\rho_{\kappa_{1}}^{n}(\mu_{n},\boldsymbol{\theta}^{\kappa_{2}})=\mu_{n}\cos\theta^{\kappa_{2}}$, $\theta^{\kappa_{2}}$ is the direction of the target and clutter, and $\boldsymbol{n}\sim\mathcal{CN}(\boldsymbol{0},N_{0}\boldsymbol{I})$ is AWGN with mean $\boldsymbol{0}$ and variance $N_{0}$.
The communication signal is denoted as
\begin{small}
\begin{align}
y=\boldsymbol{h}^{H}(\boldsymbol{x},\boldsymbol{\theta}^{T})\boldsymbol{w}x+\bar{n},\label{pro2}%
\end{align}
\end{small}%
where $\bar{n}\in\mathcal{CN}(0,N_{0}^{\prime})$ denotes the AWGN with mean $0$ and variance $N_{0}^{\prime}$, $\boldsymbol{h}(\boldsymbol{x},\boldsymbol{\theta}^{T})=\alpha\boldsymbol{a}_{T}(\boldsymbol{x},\boldsymbol{\theta}^{T})$ and $\alpha$ is the complex path gain. Based on (\ref{pro1}), the target and clutter reflected signal are processed by the combined vector $\boldsymbol{u}\in\mathbb{C}^{N_{T}\times 1}$ at the BS which are expressed as
\begin{small}
\begin{align}
&y_{r}=\boldsymbol{u}^{H}\boldsymbol{r}=\sigma\boldsymbol{u}^{H}\boldsymbol{a}_{R}(\boldsymbol{y},\boldsymbol{\theta}^{P})\boldsymbol{a}_{T}^{H}(\boldsymbol{x},\boldsymbol{\theta}^{P})\boldsymbol{w}x+\delta\boldsymbol{u}^{H}\nonumber\\
&\boldsymbol{a}_{R}(\boldsymbol{y},\boldsymbol{\theta}^{C})\boldsymbol{a}_{T}^{H}(\boldsymbol{x},\boldsymbol{\theta}^{C})\boldsymbol{w}x+\boldsymbol{u}^{H}\boldsymbol{n}.\label{pro3}
\end{align}
\end{small}%
According to \cite{10119080}, we can derive the sensing SNR using (\ref{pro3}) as
\begin{small}
\begin{align}
&\gamma_{r}=\frac{\sigma^{2}|\boldsymbol{u}^{H}\boldsymbol{A}(\boldsymbol{x},\boldsymbol{y},\boldsymbol{\theta}^{P})\boldsymbol{w}|^{2}}{\delta^{2}\boldsymbol{u}^{H}\left(\boldsymbol{A}(\boldsymbol{x},\boldsymbol{y},\boldsymbol{\theta}^{C})\boldsymbol{w}\boldsymbol{w}^{H}\boldsymbol{A}(\boldsymbol{x},\boldsymbol{y},\boldsymbol{\theta}^{C})^{H}+\boldsymbol{I}\frac{N_{0}}{\delta^{2}}\right)\boldsymbol{u}},\label{pro4}
\end{align}
\end{small}%
where $\boldsymbol{A}(\boldsymbol{x},\boldsymbol{y},\boldsymbol{\theta}^{\kappa_{2}})=\boldsymbol{a}_{R}(\boldsymbol{y},\boldsymbol{\theta}^{\kappa_{2}})\boldsymbol{a}_{T}^{H}(\boldsymbol{x},\boldsymbol{\theta}^{\kappa_{2}})$. $\gamma_{r}$ is rewritten as
\begin{small}
\begin{align}
\gamma_{r}=\frac{\sigma^{2}|\boldsymbol{u}^{H}\boldsymbol{A}(\boldsymbol{x},\boldsymbol{y},\boldsymbol{\theta}^{P})\boldsymbol{w}|^{2}}{\delta^{2}|\boldsymbol{u}^{H}\boldsymbol{A}(\boldsymbol{x},\boldsymbol{y},\boldsymbol{\theta}^{C})\boldsymbol{w}|^{2}+\frac{N_{0}}{\delta^{2}}}.\label{pro5}
\end{align}
\end{small}%
The communication SNR is expressed as
\begin{small}
\begin{align}
\gamma_{t}=\frac{|\boldsymbol{h}^{H}(\boldsymbol{x},\boldsymbol{\theta}^{T})\boldsymbol{w}|^{2}}{N_{0}^{\prime}}.\label{pro6}
\end{align}
\end{small}%
Therefore, the problem formulation is expressed as
\begin{small}
\begin{subequations}
\begin{align}
\max_{\boldsymbol{w},\boldsymbol{u},\boldsymbol{x},\boldsymbol{y}}&~\gamma_{t},\label{pro7a}\\
\mbox{s.t.}~
&\boldsymbol{x}\in \mathcal{D}_{T},&\label{pro7b}\\
&\boldsymbol{y}\in \mathcal{D}_{R},&\label{pro7c}\\
&\|x_{k}-x_{l}\|_{2}\geq D, k\neq l\in\mathcal{N}&\label{pro7d}\\
&\|y_{k}-y_{l}\|_{2}\geq D, k\neq l\in\mathcal{N}&\label{pro7e}\\
&\|\boldsymbol{w}\|\leq P,&\label{pro7f}\\
&\boldsymbol{u}^{H}\boldsymbol{u}=\|\boldsymbol{u}\|_{2}^{2}=1,&\label{pro7g}\\
&\gamma_{r}\geq \Gamma.&\label{pro7h}
\end{align}\label{pro7}%
\end{subequations}
\end{small}%
The objective of the problem described in (\ref{pro7}) is to maximize
the communication SNR for the ISAC-MA
system. (\ref{pro7d}) is the minimum distance constraints of the transmit MAs and (\ref{pro7e}) is the minimum distance constraints of the receive MAs, where $D$ is the minimum distance between antennas. (\ref{pro7f}) is the transmit power constraint and $P$ denotes the maximum transmit power. (\ref{pro7f}) is the transmit power constraint and $P$ denotes the maximum transmit power. (\ref{pro7g}) ensures that the gain for the radar target direction is set to unity. (\ref{pro7h}) is the sensing rate constraint for the radar operation that satisfies the performance requirements. Since the problem in (\ref{pro7}) is a multivariable coupling problem, the problem is non-convex. 
To solve this non-convex problem, this paper proposes a BCD algorithm based on SCA.

\section{Proposed BCD Algorithm}
\subsection{Combined beamforming optimization}
According to \cite{5706378}, we apply the minimum variance distortionless response (MVDR) weights to the radar signal in (\ref{pro4}) and $\boldsymbol{u}$ is expressed as
\begin{small}
\begin{align}
&\boldsymbol{u}=\frac{\boldsymbol{\Xi}^{-1}(\boldsymbol{w})\boldsymbol{A}(\boldsymbol{x},\boldsymbol{y},\boldsymbol{\theta}^{P})}{\boldsymbol{w}^{H}\boldsymbol{A}^{H}(\boldsymbol{x},\boldsymbol{y},\boldsymbol{\theta}^{P})\boldsymbol{\Xi}^{-1}(\boldsymbol{w})\boldsymbol{A}(\boldsymbol{x},\boldsymbol{y},\boldsymbol{\theta}^{P})\boldsymbol{w}},\label{pro8}
\end{align}
\end{small}%
where $\boldsymbol{\Xi}^{-1}(\boldsymbol{w})=\boldsymbol{A}(\boldsymbol{x},\boldsymbol{y},\boldsymbol{\theta}^{C})\boldsymbol{w}\boldsymbol{w}^{H}\boldsymbol{A}^{H}(\boldsymbol{x},\boldsymbol{y},\boldsymbol{\theta}^{C})+\boldsymbol{I}N_{0}$.

\subsection{Transmitted beamforming optimization}
Based on (\ref{pro8}), $\gamma_{r}$ is further expressed as
\begin{small}
\begin{align}
&\gamma_{r}=\sigma\boldsymbol{w}^{H}\boldsymbol{A}^{H}(\boldsymbol{x},\boldsymbol{y},\boldsymbol{\theta}^{P})(\boldsymbol{A}(\boldsymbol{x},\boldsymbol{y},\boldsymbol{\theta}^{C})\boldsymbol{w}\boldsymbol{w}^{H}\boldsymbol{A}^{H}(\boldsymbol{x},\boldsymbol{y},\boldsymbol{\theta}^{C})\nonumber\\
&+\boldsymbol{I}N_{0})^{-1}\boldsymbol{A}(\boldsymbol{x},\boldsymbol{y},\boldsymbol{\theta}^{P})\boldsymbol{w}.\label{pro9}
\end{align}
\end{small}%
Based on \cite{9650720}, $(\boldsymbol{A}(\boldsymbol{x},\boldsymbol{y},\boldsymbol{\theta}^{C})\boldsymbol{w}\boldsymbol{w}^{H}\boldsymbol{A}^{H}(\boldsymbol{x},\boldsymbol{y},\boldsymbol{\theta}^{C})+\boldsymbol{I}N_{0})^{-1}$ is expanded as
\begin{small}
\begin{align}
 \frac{1}{N_{0}}\left(\boldsymbol{I}-\frac{\boldsymbol{A}(\boldsymbol{x},\boldsymbol{y},\boldsymbol{\theta}^{C})\boldsymbol{w}\boldsymbol{w}^{H}\boldsymbol{A}^{H}(\boldsymbol{x},\boldsymbol{y},\boldsymbol{\theta}^{C})}{N_{0}+\boldsymbol{w}^{H}\boldsymbol{A}^{H}(\boldsymbol{x},\boldsymbol{y},\boldsymbol{\theta}^{C})\boldsymbol{A}(\boldsymbol{x},\boldsymbol{y},\boldsymbol{\theta}^{C})\boldsymbol{w}}\right).\label{pro10}
\end{align}
\end{small}%
$\gamma_{r}$ is rewritten as
\begin{small}
\begin{align}
&\gamma_{r}=\frac{\sigma}{N_{0}}\boldsymbol{w}^{H}\boldsymbol{A}^{H}(\boldsymbol{x},\boldsymbol{y},\boldsymbol{\theta}^{P})\nonumber\\
&\times\left(\boldsymbol{I}-\frac{\boldsymbol{A}(\boldsymbol{x},\boldsymbol{y},\boldsymbol{\theta}^{C})\boldsymbol{w}\boldsymbol{w}^{H}\boldsymbol{A}^{H}(\boldsymbol{x},\boldsymbol{y},\boldsymbol{\theta}^{C})}{N_{0}+\boldsymbol{w}^{H}\boldsymbol{A}^{H}(\boldsymbol{x},\boldsymbol{y},\boldsymbol{\theta}^{C})\boldsymbol{A}(\boldsymbol{x},\boldsymbol{y},\boldsymbol{\theta}^{C})\boldsymbol{w}}\right)\boldsymbol{A}(\boldsymbol{x},\boldsymbol{y},\boldsymbol{\theta}^{P})\boldsymbol{w}.\label{pro11}
\end{align}
\end{small}%
We let 
\begin{small}
\begin{align}
&\Psi_{1}=\mathrm{tr}(\boldsymbol{w}^{H}\boldsymbol{A}^{H}(\boldsymbol{x},\boldsymbol{y},\boldsymbol{\theta}^{P})\boldsymbol{A}(\boldsymbol{x},\boldsymbol{y},\boldsymbol{\theta}^{C})\boldsymbol{w})=\mathrm{tr}(\boldsymbol{A}_{1}\boldsymbol{W}),\Psi_{2}=\nonumber\\
&\mathrm{tr}(\boldsymbol{w}^{H}\boldsymbol{A}^{H}(\boldsymbol{x},\boldsymbol{y},\boldsymbol{\theta}^{P})\boldsymbol{A}(\boldsymbol{x},\boldsymbol{y},\boldsymbol{\theta}^{P})\boldsymbol{w})=\mathrm{tr}(\boldsymbol{A}_{2}\boldsymbol{W}),\Psi_{3}=\nonumber\\
&\mathrm{tr}(\boldsymbol{w}^{H}\boldsymbol{A}^{H}(\boldsymbol{x},\boldsymbol{y},\boldsymbol{\theta}^{C})\boldsymbol{A}(\boldsymbol{x},\boldsymbol{y},\boldsymbol{\theta}^{C})\boldsymbol{w})=\mathrm{tr}(\boldsymbol{A}_{3}\boldsymbol{W}),\label{pro12}
\end{align}
\end{small}%
in which $\boldsymbol{A}_{1}=\boldsymbol{A}^{H}(\boldsymbol{x},\boldsymbol{y},\boldsymbol{\theta}^{P})\boldsymbol{A}(\boldsymbol{x},\boldsymbol{y},\boldsymbol{\theta}^{C})$, $\boldsymbol{A}_{2}=\boldsymbol{A}^{H}(\boldsymbol{x},$ $\boldsymbol{y},\boldsymbol{\theta}^{P})\boldsymbol{A}(\boldsymbol{x},\boldsymbol{y},\boldsymbol{\theta}^{P})$ and $\boldsymbol{A}_{3}=\boldsymbol{A}^{H}(\boldsymbol{x},\boldsymbol{y},\boldsymbol{\theta}^{C})\boldsymbol{A}(\boldsymbol{x},\boldsymbol{y},\boldsymbol{\theta}^{C})$.
The optimization problem on $\boldsymbol{W}$ is expressed as
\begin{small}
\begin{subequations}
\begin{align}
\max_{\boldsymbol{W},\Psi_{1},\Psi_{2},\Psi_{3}}&~\frac{1}{N_{0}^{\prime}}\mathrm{tr}(\boldsymbol{H}\boldsymbol{W}),\label{pro13a}\\
\mbox{s.t.}~
&\mathrm{tr}(\boldsymbol{W})\leq P_{t},&\label{pro13b}\\
&\Psi_{2}>0,\Psi_{3}>0,&\label{pro13c}\\
&\Psi_{i}=\mathrm{tr}(\boldsymbol{A}_{i}\boldsymbol{W}), i\in\{1,2,3\},&\label{pro13d}\\
&\frac{\sigma}{N_{0}}\left(\Psi_{2}-\frac{|\Psi_{1}|^{2}}{N_{0}+\Psi_{3}}\right)>\Gamma,&\label{pro13e}\\
&\mathrm{rank}(\boldsymbol{W})=1,&\label{pro13f}
\end{align}\label{pro13}%
\end{subequations}
\end{small}%
in which $\boldsymbol{H}=\boldsymbol{h}(\boldsymbol{x},\boldsymbol{\theta}^{T})\boldsymbol{h}^{H}(\boldsymbol{x},\boldsymbol{\theta}^{T})$. The objective function in (\ref{pro13a}) and constraints in (\ref{pro13b})-(\ref{pro13d}) are convex. Next, we deal with the constraint condition in (\ref{pro13e}).
\begin{small}
\begin{align}
\frac{\sigma}{N_{0}}(N_{0}+\Psi_{3})\Psi_{2}-|\Psi_{1}|^{2}>\Gamma(N_{0}+\Psi_{3}).\label{pro14}
\end{align}
\end{small}%
Since $\Psi_{2}\Psi_{3}$ is coupled, the constraint condition is still non-convex. We let $\Omega=\Psi_{2}\Psi_{3}$, the constraint condition in (\ref{pro13e}) is rewritten as
\begin{small}
\begin{align}
\frac{\sigma}{N_{0}}\Psi_{2}N_{0}+\Omega-|\Psi_{1}|^{2}>\Gamma(N_{0}+\Psi_{3}).\label{pro15}
\end{align}
\end{small}%
Constraint (\ref{pro15}) is convex, therefore, problem (\ref{pro13}) is rewritten as
\begin{small}
\begin{subequations}
\begin{align}
\max_{\boldsymbol{W},\Psi_{1},\Psi_{2},\Psi_{3},\Omega}&~\frac{1}{N_{0}^{\prime}}\mathrm{tr}(\boldsymbol{H}\boldsymbol{W}),\label{pro16a}\\
\mbox{s.t.}~
&(\ref{pro13b}),(\ref{pro13c}),(\ref{pro13d}),(\ref{pro13f}),&\label{pro16b}\\
&\Omega=\Psi_{2}\Psi_{3},\Omega>0, &\label{pro16c}\\
&\frac{\sigma}{N_{0}}\Psi_{2}N_{0}+\Omega-|\Psi_{1}|^{2}>\Gamma(N_{0}+\Psi_{3}).&\label{pro16d}
\end{align}\label{pro16}%
\end{subequations}
\end{small}%
To deal with the non-convex constraint (\ref{pro16c}). The equality constraint in (\ref{pro16c}) can be re-expressed as
\begin{small}
\begin{align}
&\Psi_{2}\Psi_{3}\geq \Omega,
\Psi_{2}\Psi_{3}\leq \Omega.\label{pro17}
\end{align}
\end{small}%
We use the Taylor series and inequality of arithmetic and geometric means(IA-GM) to deal with non-convex constraint (\ref{pro17}), they are given by
\begin{small}
\begin{align}
&\frac{1}{\Psi_{2}\Psi_{3}}\leq \frac{1}{\Omega},~\Psi_{2}\Psi_{3}\leq\frac{\Psi_{2}^{2}+\Psi_{3}^{2}}{2}\leq\Omega.\label{pro18}
\end{align}
\end{small}%
Since $\frac{1}{\Psi_{2}\Psi_{3}}\leq \frac{1}{\Omega}$ is non-convex, we use SCA based on the Taylor series to obtain the lower bound of $\frac{1}{\Omega}$ is rewritten as
\begin{small}
\begin{align}
&\frac{1}{\Psi_{2}\Psi_{3}}\leq \frac{1}{\Omega_{0}}-\frac{1}{\Omega_{0}^{2}}(\Omega-\Omega_{0}).\label{pro19}
\end{align}
\end{small}%
Problem (\ref{pro16}) is rewritten as
\begin{small}
\begin{subequations}
\begin{align}
\max_{\boldsymbol{W},\Psi_{1},\Psi_{2}, \Psi_{3},\Omega}&~\frac{1}{N_{0}^{\prime}}\mathrm{tr}(\boldsymbol{H}\boldsymbol{W}),\label{pro20a}\\
\mbox{s.t.}~
&(\ref{pro13b}),(\ref{pro13c}),(\ref{pro13d}),(\ref{pro13f}),(\ref{pro16d}), &\label{pro20b}\\
&\frac{1}{\Psi_{2}\Psi_{3}}\leq \frac{1}{\Omega_{0}}-\frac{1}{\Omega_{0}^{2}}(\Omega-\Omega_{0}),&\label{pro20c}\\
&\frac{\Psi_{2}^{2}+\Psi_{3}^{2}}{2}\leq\Omega.&\label{pro20d}
\end{align}\label{pro20}%
\end{subequations}
\end{small}%
Problem (\ref{pro20}) is non-convex due to the non-convex constraint in (\ref{pro13f}). To deal with (\ref{pro13f}), we use the SDR method, thus problem (\ref{pro20}) is rewritten as
\begin{small}
\begin{subequations}
\begin{align}
\max_{\boldsymbol{W},\Psi_{1},\Psi_{2}, \Psi_{3},\Omega}&~\frac{1}{N_{0}^{\prime}}\mathrm{tr}(\boldsymbol{H}\boldsymbol{W}),\label{pro21a}\\
\mbox{s.t.}~
&(\ref{pro13b}),(\ref{pro13c}),(\ref{pro13d}),(\ref{pro13f}),(\ref{pro16d}), (\ref{pro20c}),(\ref{pro20d}).&\label{pro21b}
\end{align}\label{pro21}%
\end{subequations}
\end{small}%
We can now use CVX to solve (\ref{pro21}) which is convex. The optimal solution $\boldsymbol{W}^{*}$ to  (\ref{pro21}) always satisfies that $\mathrm{rank}(\boldsymbol{W}^{*}
) = 1$, the proof details refer to \cite{9930110}.

\subsection{Position optimization of MAs}
Given $\boldsymbol{w}$, $\boldsymbol{u}$ and $\boldsymbol{y}$, problem (\ref{pro7}) is rewritten as
\begin{small}
\begin{subequations}
\begin{align}
\max_{\boldsymbol{x}}&~\frac{|\boldsymbol{h}^{H}(\boldsymbol{x},\boldsymbol{\theta}^{T})\boldsymbol{w}|^{2}}{N_{0}^{\prime}},\label{pro22a}\\
\mbox{s.t.}~
&(\ref{pro7b}),(\ref{pro7d}),&\label{pro22b}\\
&\frac{\sigma}{N_{0}}\boldsymbol{w}^{H}\boldsymbol{A}^{H}(\boldsymbol{x},\boldsymbol{y},\boldsymbol{\theta}^{P})&\nonumber\\
&\times\left(\boldsymbol{I}-\frac{\boldsymbol{A}(\boldsymbol{x},\boldsymbol{y},\boldsymbol{\theta}^{C})\boldsymbol{w}\boldsymbol{w}^{H}\boldsymbol{A}^{H}(\boldsymbol{x},\boldsymbol{y},\boldsymbol{\theta}^{C})}{N_{0}+\boldsymbol{w}^{H}\boldsymbol{A}^{H}(\boldsymbol{x},\boldsymbol{y},\boldsymbol{\theta}^{C})\boldsymbol{A}(\boldsymbol{x},\boldsymbol{y},\boldsymbol{\theta}^{C})\boldsymbol{w}}\right)&\nonumber\\
&\times\boldsymbol{A}(\boldsymbol{x},\boldsymbol{y},\boldsymbol{\theta}^{P})\boldsymbol{w}\geq \Gamma.&\label{pro22c}
\end{align}\label{pro22}%
\end{subequations}
\end{small}
We introduce slack variables $\Phi$, $\bar{\Phi}$ and $\tilde{\Phi}$ and have
\begin{small}
\begin{align}
&\boldsymbol{w}^{H}\boldsymbol{A}^{H}(\boldsymbol{x},\boldsymbol{y},\boldsymbol{\theta}^{P})\boldsymbol{A}(\boldsymbol{x},\boldsymbol{y},\boldsymbol{\theta}^{P})\boldsymbol{w}\leq \Phi,\nonumber\\
&\boldsymbol{w}^{H}\boldsymbol{A}^{H}(\boldsymbol{x},\boldsymbol{y},\boldsymbol{\theta}^{C})\boldsymbol{A}(\boldsymbol{x},\boldsymbol{y},\boldsymbol{\theta}^{C})\boldsymbol{w}\leq \bar{\Phi},\nonumber\\
&|\boldsymbol{w}^{H}\boldsymbol{A}^{H}(\boldsymbol{x},\boldsymbol{y},\boldsymbol{\theta}^{P})\boldsymbol{A}(\boldsymbol{x},\boldsymbol{y},\boldsymbol{\theta}^{C})\boldsymbol{w}|^{2}\geq \tilde{\Phi}.\label{pro23}
\end{align}
\end{small}
Problem (\ref{pro22}) is rewritten as
\begin{small}
\begin{subequations}
\begin{align}
\max_{\boldsymbol{x},\Phi, \bar{\Phi},\tilde{\Phi}}&~\frac{|\boldsymbol{h}^{H}(\boldsymbol{x},\theta_{t})\boldsymbol{w}|^{2}}{N_{0}^{\prime}},\label{pro24a}\\
\mbox{s.t.}~
&(\ref{pro7b}),(\ref{pro7d}),&\label{pro24b}\\
&\frac{\sigma}{N_{0}}\left(\Phi-\frac{\tilde{\Phi}}{N_{0}+\bar{\Phi}}\right)\geq \Gamma,&\label{pro24c}\\
&(\ref{pro23}),\Phi>0, \bar{\Phi}>0.&\label{pro24d}
\end{align}\label{pro24}%
\end{subequations}
\end{small}
According to \cite{10158711}, (\ref{pro23}) are re-expressed as
\begin{small}
\begin{align}
&N_{T}\boldsymbol{a}_{T}^{H}(\boldsymbol{x},\boldsymbol{\theta}^{P})\boldsymbol{W}\boldsymbol{a}_{T}(\boldsymbol{x},\boldsymbol{\theta}^{P}), N_{T}\boldsymbol{a}_{T}^{H}(\boldsymbol{x},\boldsymbol{\theta}^{C})\boldsymbol{W}\boldsymbol{a}_{T}(\boldsymbol{x},\boldsymbol{\theta}^{C}),\nonumber\\
&\tilde{N}\boldsymbol{a}_{T}^{H}(\boldsymbol{x},\boldsymbol{\theta}^{C})\boldsymbol{W}\boldsymbol{a}_{T}(\boldsymbol{x},\boldsymbol{\theta}^{P}),\label{pro25}
\end{align}
\end{small}%
in which $\tilde{N}=\boldsymbol{a}_{R}^{H}(\boldsymbol{y},\boldsymbol{\theta}^{P})\boldsymbol{a}_{T}(\boldsymbol{y},\boldsymbol{\theta}^{C})$, $\boldsymbol{W}=\boldsymbol{w}\boldsymbol{w}^{H}$.
We use the Taylor series to approximate (\ref{pro25}) and the objective function in (\ref{pro24a}) and have
\begin{small}
\begin{align}
&\frac{|\boldsymbol{h}^{H}(\boldsymbol{x},\boldsymbol{\theta}^{T})\boldsymbol{w}|^{2}}{N_{0}^{\prime}}\approx 2\frac{1}{N_{0}^{\prime}}\mathrm{Re}\{\boldsymbol{a}_{T}^{H}(\boldsymbol{x}^{i},\boldsymbol{\theta}^{T})\boldsymbol{W}\boldsymbol{a}_{T}(\boldsymbol{x},\boldsymbol{\theta}^{T})\}-\nonumber\\
&\frac{1}{N_{0}^{\prime}}\boldsymbol{a}_{T}^{H}(\boldsymbol{x}^{i},\boldsymbol{\theta}^{T})\boldsymbol{W}\boldsymbol{a}_{T}(\boldsymbol{x}^{i},\boldsymbol{\theta}^{T}),\label{pro_26}\\
&N_{T}\boldsymbol{a}_{T}^{H}(\boldsymbol{x},\boldsymbol{\theta}^{P})\boldsymbol{W}\boldsymbol{a}_{T}(\boldsymbol{x},\boldsymbol{\theta}^{P})\approx 2N_{T}\mathrm{Re}\{\boldsymbol{a}_{T}^{H}(\boldsymbol{x}^{i},\boldsymbol{\theta}^{P})\boldsymbol{W}\boldsymbol{a}_{T}(\boldsymbol{x},\boldsymbol{\theta}^{P})\}\nonumber\\
&-N_{T}\boldsymbol{a}_{T}^{H}(\boldsymbol{x}^{i},\boldsymbol{\theta}^{P})\boldsymbol{W}\boldsymbol{a}_{T}(\boldsymbol{x}^{i},\boldsymbol{\theta}^{P}),\nonumber\\
&N_{T}\boldsymbol{a}_{T}^{H}(\boldsymbol{x},\boldsymbol{\theta}^{C})\boldsymbol{W}\boldsymbol{a}_{T}(\boldsymbol{x},\boldsymbol{\theta}^{C})\approx 2N_{T}\mathrm{Re}\{\boldsymbol{a}_{T}^{H}(\boldsymbol{x}^{i},\boldsymbol{\theta}^{C})\boldsymbol{W}\boldsymbol{a}_{T}(\boldsymbol{x},\boldsymbol{\theta}^{C})\}\nonumber\\
&-N_{T}\boldsymbol{a}_{T}^{H}(\boldsymbol{x}^{i},\boldsymbol{\theta}^{C})\boldsymbol{W}\boldsymbol{a}_{T}(\boldsymbol{x}^{i},\boldsymbol{\theta}^{C}),\label{pro__26}\\
&\tilde{N}\boldsymbol{a}_{T}^{H}(\boldsymbol{x},\boldsymbol{\theta}^{C})\boldsymbol{W}\boldsymbol{a}_{T}(\boldsymbol{x},\boldsymbol{\theta}^{P})\approx \tilde{N}\boldsymbol{a}_{T}^{H}(\boldsymbol{x}^{i},\boldsymbol{\theta}^{P})\boldsymbol{W}\boldsymbol{a}_{T}(\boldsymbol{x},\boldsymbol{\theta}^{C})+\tilde{N}\nonumber\\
&\boldsymbol{a}_{T}^{H}(\boldsymbol{x}^{i},\boldsymbol{\theta}^{C})\boldsymbol{W}\boldsymbol{a}_{T}(\boldsymbol{x},\boldsymbol{\theta}^{P})-\tilde{N}\boldsymbol{a}_{T}^{H}(\boldsymbol{x}^{i},\boldsymbol{\theta}^{P})\boldsymbol{W}\boldsymbol{a}_{T}(\boldsymbol{x}^{i},\boldsymbol{\theta}^{C}).\label{pro26}
\end{align}
\end{small}%
It is not difficult to find (\ref{pro_26}) and (\ref{pro__26}) are still non-convex. To deal with these non-convex function, we continue to approximate (\ref{pro__26}). We assume $\boldsymbol{b}^{T}=\boldsymbol{a}_{T}^{H}(\boldsymbol{x}^{i},\boldsymbol{\theta}^{T})$, $\boldsymbol{b}^{P}=\boldsymbol{a}_{T}^{H}(\boldsymbol{x}^{i},\boldsymbol{\theta}^{P})$ and $\boldsymbol{b}^{C}=\boldsymbol{a}_{T}^{H}(\boldsymbol{x}^{i},\boldsymbol{\theta}^{C})$, we have
\begin{small}
\begin{align}
&\mathrm{Re}\{\boldsymbol{b}^{T}\boldsymbol{a}_{T}(\boldsymbol{x}^{i},\boldsymbol{\theta}^{T})\}=\sum\nolimits_{i=1}^{N_{T}}|\boldsymbol{b}^{T}_{i}|\cos(\rho_{T}^{i}(x_{i},\theta_{i}^{T})-\angle\boldsymbol{b}^{T}_{i})\nonumber\\
&\mathrm{Re}\{\boldsymbol{b}^{P}\boldsymbol{a}_{T}(\boldsymbol{x}^{i},\boldsymbol{\theta}^{P})\}=\sum\nolimits_{i=1}^{N_{T}}|\boldsymbol{b}^{P}_{i}|\cos(\rho_{T}^{i}(x_{i},\theta_{i}^{P})-\angle\boldsymbol{b}^{P}_{i})\nonumber\\
&\mathrm{Re}\{\boldsymbol{b}^{C}\boldsymbol{a}_{T}(\boldsymbol{x}^{i},\boldsymbol{\theta}^{C})\}=\sum\nolimits_{i=1}^{N_{T}}|\boldsymbol{b}^{C}_{i}|\cos(\rho_{T}^{i}(x_{i},\theta_{i}^{C})-\angle\boldsymbol{b}^{C}_{i}).\label{pro27}
\end{align}
\end{small}%
We use the Taylor series $\cos(x)\approx 1-\frac{x^{2}}{2}$ to approximate $\cos(\rho_{T}^{i}(x_{i},\theta_{i}^{T})-\angle\boldsymbol{b}^{T}_{i})$, $\cos(\rho_{T}^{i}(x_{i},\theta_{i}^{P})-\angle\boldsymbol{b}^{P}_{i})$ and $\cos(\rho_{T}^{i}(x_{i},\theta_{i}^{C})-\angle\boldsymbol{b}^{C}_{i})$ and they are expressed as
\begin{small}
\begin{align}
&\cos(\rho_{T}^{i}(x_{i},\theta_{i}^{T})-\angle\boldsymbol{b}^{T}_{i})\approx 1-(\rho_{T}^{i}(x_{i},\theta_{i}^{T})-\angle\boldsymbol{b}^{T}_{i})^{2}/2,\\
&\cos(\rho_{T}^{i}(x_{i},\theta_{i}^{P})-\angle\boldsymbol{b}^{P}_{i})\approx 1-(\rho_{T}^{i}(x_{i},\theta_{i}^{P})-\angle\boldsymbol{b}^{P}_{i})^{2}/2,\nonumber\\
&\cos(\rho_{T}^{i}(x_{i},\theta_{i}^{C})-\angle\boldsymbol{b}^{C}_{i})\approx 1-(\rho_{T}^{i}(x_{i},\theta_{i}^{C})-\angle\boldsymbol{b}^{C}_{i})^{2}/2.\label{pro28}
\end{align}
\end{small}%
Then, 
we let $\hat{\lambda}_{i}=\rho_{T}^{i}(x_{i},\theta_{i}^{T})$, $\bar{\kappa}_{i}=\rho_{P}^{i}(x_{i},\theta_{i}^{P})$, $\kappa_{i}=\rho_{C}^{i}(x_{i},\theta_{i}^{C})$ and $\tau^{T}_{i}=(\hat{\lambda}_{i}-\angle\boldsymbol{b}^{T}_{i})^{2}$, $\tau^{P}_{i}=(\bar{\kappa}_{i}-\angle\boldsymbol{b}^{P}_{i})^{2}$ and $\tau^{C}_{i}=(\kappa_{i}-\angle\boldsymbol{b}^{C}_{i})^{2}$. Thus (\ref{pro27}) can be approximated as
\begin{small}
\begin{align}
&2/N_{0}^{\prime}\sum\nolimits_{i=1}^{N_{T}}|\boldsymbol{b}^{T}_{i}|\left(1-\frac{(\tau^{T}_{i})^{2}}{2}\right)-N_{T}\boldsymbol{a}_{T}^{H}(\boldsymbol{x}^{i},\boldsymbol{\theta}^{T})\boldsymbol{W}\boldsymbol{a}_{T}(\boldsymbol{x}^{i},\boldsymbol{\theta}^{T}),\nonumber\\
&2N_{T}\sum\nolimits_{i=1}^{N_{T}}|\boldsymbol{b}^{P}_{i}|\left(1-\frac{(\tau^{P}_{i})^{2}}{2}\right)-N_{T}\boldsymbol{a}_{T}^{H}(\boldsymbol{x}^{i},\boldsymbol{\theta}^{P})\boldsymbol{W}\boldsymbol{a}_{T}(\boldsymbol{x}^{i},\boldsymbol{\theta}^{P}),\nonumber\\
&2N_{T}\sum\nolimits_{i=1}^{N_{T}}|\boldsymbol{b}^{C}_{i}|\left(1-\frac{(\tau^{C}_{i})^{2}}{2}\right)-N_{T}\boldsymbol{a}_{T}^{H}(\boldsymbol{x}^{i},\boldsymbol{\theta}^{C})\boldsymbol{W}\boldsymbol{a}_{T}(\boldsymbol{x}^{i},\boldsymbol{\theta}^{C}), \label{pro__29}
\end{align}
\end{small}%
and (\ref{pro26}) is rewritten as
\begin{small}
\begin{align}
&\tilde{N}\boldsymbol{b}^{P}\boldsymbol{a}_{T}(\boldsymbol{x},\boldsymbol{\theta}^{C})+\tilde{N}\boldsymbol{b}^{C}\boldsymbol{a}_{T}(\boldsymbol{x},\boldsymbol{\theta}^{P})-\tilde{N}\boldsymbol{a}_{T}^{H}(\boldsymbol{x}^{i},\boldsymbol{\theta}^{P})\nonumber\\
&\times\boldsymbol{W}\boldsymbol{a}_{T}(\boldsymbol{x}^{i},\boldsymbol{\theta}^{C}).\label{pro29}
\end{align}
\end{small}%
We continue to deal with (\ref{pro29}), and $\tilde{N}\boldsymbol{b}^{P}\boldsymbol{a}_{T}(\boldsymbol{x},\boldsymbol{\theta}^{C})+\tilde{N}\boldsymbol{b}^{C}\boldsymbol{a}_{T}(\boldsymbol{x},\boldsymbol{\theta}^{P})$ is rewritten as (\ref{pro_29}) on the top of page, where $\Omega_{i}^{P}=\tilde{N}|\boldsymbol{b}_{i}^{P}|\cos(\angle\boldsymbol{b}^{P}_{i})-\tilde{N}j|\boldsymbol{b}_{i}^{P}|\sin(\angle\boldsymbol{b}^{P}_{i})$,~$\Xi_{i}^{P}=\tilde{N}|\boldsymbol{b}_{i}^{P}|\cos(\angle\boldsymbol{b}^{P}_{i})+\tilde{N}j|\boldsymbol{b}_{i}^{P}|\sin(\angle\boldsymbol{b}^{P}_{i})$, $\bar{\Omega}_{i}^{C}=\tilde{N}|\boldsymbol{b}_{i}^{C}|\cos(\angle\boldsymbol{b}^{C}_{i})$\\
$-\tilde{N}j|\boldsymbol{b}_{i}^{C}|\sin(\angle\boldsymbol{b}^{C}_{i})$, $\bar{\Xi}_{i}^{C}=\tilde{N}|\boldsymbol{b}_{i}^{C}|\cos(\angle\boldsymbol{b}^{C}_{i})+j\tilde{N}|\boldsymbol{b}_{i}^{C}|\sin$\\
$(\angle\boldsymbol{b}^{C}_{i})$.

\begin{figure*}
\begin{small}
\begin{align}
&(\ref{pro29})=\sum\nolimits_{i=1}^{N_{T}}\Omega_{i}^{P}\cos(\rho_{T}^{i}(x_{i},\theta_{i}^{C}))+\Xi_{i}^{P}\sin(\rho_{T}^{i}(x_{i},\theta_{i}^{C}))+\sum\nolimits_{i=1}^{N_{T}}\bar{\Omega}_{i}^{C}\cos(\rho_{T}^{i}(x_{i},\theta_{i}^{P}))+\bar{\Xi}_{i}^{C}\sin(\rho_{T}^{i}(x_{i},\theta_{i}^{P}))\nonumber\\
&\approx\sum\nolimits_{i=1}^{N_{T}}\Omega_{i}^{P}\left(1-\kappa_{i}^{2}/2\right)+\Xi_{i}^{P}\left(\kappa_{i}-\kappa_{i}^{3}/6\right)+\bar{\Omega}_{i}^{C}\left(1-\bar{\kappa}_{i}^{2}/2\right)+\bar{\Xi}_{i}^{C}\left(\bar{\kappa}_{i}-\bar{\kappa}_{i}^{3}/6\right).\label{pro_29}
\end{align}
\end{small}
\hrulefill
\end{figure*}
 We introduce auxiliary variable $t$ and let $\eta$=(\ref{pro_29}). To deal with the higher order variables in (\ref{pro_29}), we continue to introduce the following variables
\begin{small}
\begin{align}
&\kappa_{i}^{2}=u_{i}, \zeta_{i}=\kappa_{i}u_{i}, \bar{\kappa}_{i}^{2}=\bar{u}_{i},\bar{\zeta}_{i}=\bar{\kappa}_{i}\bar{u}_{i}, \tau^{T}_{i}=(\hat{\lambda}_{i}-\angle\boldsymbol{b}^{T}_{i})^{2},\nonumber\\
&\tau^{P}_{i}=(\bar{\kappa}_{i}-\angle\boldsymbol{b}^{P}_{i})^{2},\tau^{C}_{i}=(\kappa_{i}-\angle\boldsymbol{b}^{C}_{i})^{2}.\label{pro_34}
\end{align}    
\end{small}%

\begin{figure*}
\begin{small}
\begin{align}
&\eta=\sum\nolimits_{i=1}^{N_{T}}\Omega_{i}^{P}\left(1-u_{i}/2\right)+\Xi_{i}^{P}\left(\kappa_{i}-\zeta_{i}/6\right)+\bar{\Omega}_{i}^{C}\left(1-\bar{u}_{i}/2\right)+\bar{\Xi}_{i}^{C}\left(\bar{\kappa}_{i}-\bar{\zeta}_{i}/6\right).\label{pro_30}
\end{align}
\end{small}
\hrulefill
\end{figure*}
According to (\ref{pro_34}), (\ref{pro_29}) is re-expressed as (\ref{pro_30}). Thus, problem (\ref{pro24}) is re-expressed as
\begin{small}
\begin{subequations}
\begin{align}
\max_{\boldsymbol{x},\{\kappa_{i},\bar{\kappa}_{i}\},\{u_{i},\bar{u}_{i}\}, \{\zeta_{i},\bar{\zeta}_{i}\},\atop\eta,\{\hat{\lambda}_{i}\},\{\tau_{i}^{T},\tau_{i}^{P},\tau_{i}^{C}\}}&~2/N_{0}^{\prime}\sum\nolimits_{i=1}^{N_{T}}|\boldsymbol{b}^{T}_{i}|\left(1-(\tau^{T}_{i})^{2}/2\right),\label{pro301a}\\
\mbox{s.t.}~
&(\ref{pro7b}),(\ref{pro7d}),(\ref{pro7e}),(\ref{pro24c}), (\ref{pro_30}),&\label{pro301b}\\
&\kappa_{i}>0,u_{i}>0,
\zeta_{i}>0,
\bar{\kappa}_{i}>0,\bar{u}_{i}>0,
&\nonumber\\
&\bar{\zeta}_{i}>0,
\hat{\lambda}_{i}>0,&\label{pro301c}\\
&|\eta|^{2}\geq \tilde{\Phi},&\label{pro301d}\\
&\kappa_{i}=\rho_{T}^{i}(x_{i},\theta_{i}^{C}),&\\
&\bar{\kappa}_{i}=\rho_{T}^{i}(x_{i},\theta_{i}^{P}).&\label{pro301e}
\end{align}\label{pro301}%
\end{subequations}
\end{small}%
Next, we use the Taylor series and inequality of arithmetic and geometric means to deal with non-convex equality constraints in (\ref{pro_34}) and based on (\ref{pro18}) and (\ref{pro19}), the variables in (\ref{pro_34}) are transformed using inequalities which are shown as (\ref{pro36}) on the top of this page. 
\begin{figure*}
\begin{small}
\begin{align}
&\kappa_{i}^{2}\leq u_{i}\leq\kappa_{i,0}^{2}+2\kappa_{i,0}(\kappa_{i}-\kappa_{i,0}), \zeta_{i}\geq (\kappa_{i}^{2}+u_{i}^{2})/2, 1/(\kappa_{i}u_{i})\leq 1/\zeta_{i,0}-1/{\zeta_{i,0}^{2}}(\zeta_{i}-\zeta_{i,0}),\nonumber\\
&\bar{\kappa}_{i}^{2}\leq  \bar{u}_{i}\leq\bar{\kappa}_{i,0}^{2}+2\bar{\kappa}_{i,0}(\bar{\kappa}_{i}-\bar{\kappa}_{i,0}), \bar{\zeta}_{i}\geq (\bar{\kappa}_{i}^{2}+\bar{u}_{i}^{2})/2, 1/(\bar{\kappa}_{i}\bar{u}_{i})\leq 1/\bar{\zeta}_{i,0}-1/{\bar{\zeta}_{i,0}^{2}}(\bar{\zeta}_{i}-\bar{\zeta}_{i,0}),\nonumber\\
&(\hat{\lambda}_{i}-\angle\boldsymbol{b}^{T}_{i})^{2}\leq 
\tau^{T}_{i}\leq(\hat{\lambda}_{i,0}-\angle\boldsymbol{b}^{T}_{i})^{2}+(\hat{\lambda}_{i,0}-\angle\boldsymbol{b}^{T}_{i})(\hat{\lambda}_{i}-\hat{\lambda}_{i,0}), (\kappa_{i,0}-\angle\boldsymbol{b}^{P}_{i})^{2}\leq 
\tau^{P}_{i}\leq(\bar{\kappa}_{i,0}-\angle\boldsymbol{b}^{P}_{i})^{2}+(\bar{\kappa}_{i,0}-\angle\boldsymbol{b}^{P}_{i})(\bar{\kappa}_{i}-\bar{\kappa}_{i,0}),\nonumber\\
&(\kappa_{i}-\angle\boldsymbol{b}^{C}_{i})^{2}\leq 
\tau^{C}_{i}\leq(\kappa_{i,0}-\angle\boldsymbol{b}^{C}_{i})^{2}+(\kappa_{i,0}-\angle\boldsymbol{b}^{C}_{i})(\kappa_{i}-\kappa_{i,0}).\label{pro36}
\end{align}
\end{small}
\hrulefill
\end{figure*}
Finally, the objective function in (\ref{pro301d}) is still non-convex and the Taylor series is used to obtain the lower bound of $|\eta|^{2}$
\begin{small}
\begin{align}
&\mathrm{Re}\{\eta_{0}^{H}\eta\}-|\eta_{0}^{2}|\geq\bar{\Phi}.\label{pro32} 
\end{align}
\end{small}%
Problem (\ref{pro301}) is rewritten as
\begin{small}
\begin{subequations}
\begin{align}
\max_{\boldsymbol{x},\{\kappa_{i},\bar{\kappa}_{i}\},\{u_{i},\bar{u}_{i}\}, \{\zeta_{i},\bar{\zeta}_{i}\},\atop\eta,\{\hat{\lambda}_{i}\},\{\tau_{i}^{T},\tau_{i}^{P},\tau_{i}^{C}\}}&~2/N_{0}^{\prime}\sum\nolimits_{i=1}^{N_{T}}|\boldsymbol{b}^{T}_{i}|\left(1-(\tau^{T}_{i})^{2}/2\right),\label{pro302a}\\
\mbox{s.t.}~
&(\ref{pro7b}),(\ref{pro7d}),(\ref{pro7e}),(\ref{pro24c}), (\ref{pro_30}),(\ref{pro301c}),&\nonumber\\
&(\ref{pro301e}),(\ref{pro32}).&\label{pro302b}
\end{align}\label{pro302}%
\end{subequations}
\end{small}%
We note that (\ref{pro7d}) and (\ref{pro7e}) are now the only non-convex constraints in (\ref{pro302}) only non-convex constraints. To deal with non-convex constraint (\ref{pro7d}), we propose the following solution. Denote the gradient of $\|x_{k}-x_{l}\|_{2}$ over $x_{k}$ as $\frac{\partial \|x_{k}-x_{l}\|_{2}}{\partial x_{k}}=(x_{k}-x_{l})/\|x_{k}-x_{l}\|_{2}$. Since the denominator
term $\|x_{k}-x_{l}\|_{2}\geq D>0$, the gradient vector always exists. As $\|x_{k}-x_{l}\|_{2}$ is a convex function with respect to $x_{k}$, we have the following inequality by applying the first-order Taylor
expansion at the given $t$-th iteration point $x_{k}^{(t)}$ and we have
\begin{small}
\begin{align}
&\|x_{k}-x_{l}\|_{2}\geq \|x_{k}^{(t)}-x_{l}\|_{2}+\left(\frac{\partial \|x_{k}-x_{l}\|_{2}}{x_{k}}\right)\Bigg|_{x_{k}=x_{k}^{(t)}}(x_{k}-x_{k}^{(t)})\nonumber\\
&=\|x_{k}^{(t)}-x_{l}\|_{2}+\frac{1}{\|x_{k}^{(t)}-x_{l}\|_{2}}(x_{k}^{(t)}-x_{l})(x_{k}-x_{k}^{(t)})\nonumber\\
&=\frac{1}{\|x_{k}^{(t)}-x_{l}\|_{2}}(x_{k}^{(t)}-x_{l})(x_{k}-x_{l}).\label{pro34}
\end{align}
\end{small}%
Therefore, (\ref{pro7d}) can be equivalently written as
\begin{small}
\begin{align}
\frac{1}{\|x_{k}^{(t)}-x_{l}\|_{2}}(x_{k}^{(t)}-x_{l})(x_{k}-x_{l})\geq D.\label{pro_314}
\end{align}
\end{small}%
(\ref{pro_314}) is linear with respect to $x_{k}$ and it is convex over $x_{k}$.
Finally the convex problem optimization problem is given by
\begin{small}
\begin{subequations}
\begin{align}
\max_{\boldsymbol{x},u_{i},\tau_{i},
v_{i},
\zeta_{i},
\xi_{i},\atop\{\bar{u}_{i},\bar{\tau}_{i},\bar{v}_{i},\bar{\zeta}_{i},\bar{\xi}_{i}\}}&~(\ref{pro302a}),\label{pro35a}\\
\mbox{s.t.}~
&(\ref{pro7d}),(\ref{pro24c}), (\ref{pro_30}),(\ref{pro301c}),(\ref{pro301e}),&\nonumber\\
&(\ref{pro32}),(\ref{pro_314}),&\label{pro35b}
\end{align}\label{pro35}%
\end{subequations}
\end{small}%
Since $\boldsymbol{y}$ is similar to $\boldsymbol{x}$, thus we can also use \textcolor{blue}{(\ref{pro34})} to evaluate the optimal of $\boldsymbol{y}$. The BCD algorithm based on SCA is summarized as follows:
\begin{algorithm}%
\caption{Proposed BCD Algorithm for Problem (\ref{pro7})} \label{algo1}
\hspace*{0.02in}{\bf Initialize:}
$\boldsymbol{w}^{(0)}$, $\boldsymbol{u}^{(0)}$, $\boldsymbol{x}^{(0)}$, $\boldsymbol{y}^{(0)}$.\\
\hspace*{0.02in}{\bf Repeat:}~$t=t+1$.\\
Given $\boldsymbol{w}^{(t)}$, $\boldsymbol{x}^{(t)}$, $\boldsymbol{y}^{(t)}$, utilizing (\ref{pro8}) to computing $\boldsymbol{u}^{(t+1)}$;\\
Given $\boldsymbol{u}^{(t+1)}$, $\boldsymbol{x}^{(t)}$, $\boldsymbol{y}^{(t)}$, utilizing (\ref{pro21}) to computing $\boldsymbol{w}^{(t+1)}$;\\
Given $\boldsymbol{u}^{(t+1)}$, $\boldsymbol{w}^{(t+1)}$, utilizing (\ref{pro35}) to computing $\boldsymbol{x}^{(t+1)}$ and $\boldsymbol{y}^{(t+1)}$;\\
\hspace*{0.02in}{\bf Until:}~$|\gamma_{t}^{t+1}-\gamma_{t}^{t}|\leq \epsilon$.\\
\hspace*{0.02in}{\bf Output:}
$\boldsymbol{w}^{(t+1)}$, $\boldsymbol{u}^{(t+1)}$, $\boldsymbol{x}^{(t+1)}$, $\boldsymbol{y}^{(t+1)}$.\\
\end{algorithm}

\section{Numerical Results}\label{V}

In the simulation, we
consider $N_{T}=N_{R}=8$ transmit MAs and receive MAs. The minimum distance between MAs is set as $D = \lambda/2$. The average SNR is deﬁned as $\frac{P}{\sigma^{2}} = 100$. Constant
values of clutter and target reflection factors are set as
$\sigma_{t}=\sigma_{c}=1$ during the simulations. The direction of the target was set towards $30$, while the clutter direction was
set as $60$. PICOS\cite{10318068} was used to specify and solve the
convex optimization problems in the simulation.

\begin{figure*}[t]
\centering
\subfigure{
\begin{minipage}[t]{0.35\linewidth}
\centering
\includegraphics[scale=0.35]{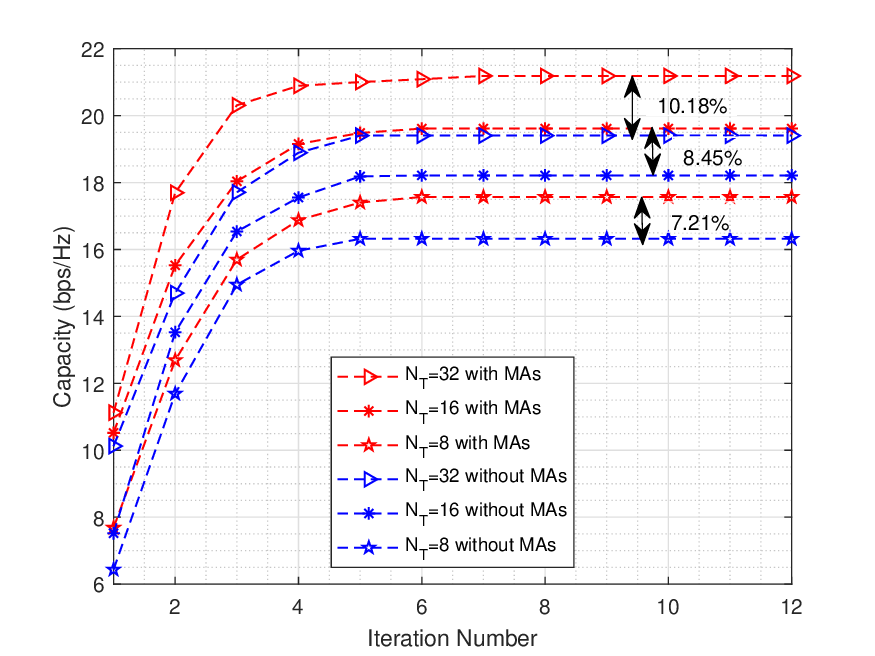}
\caption{Convergence of proposed algorithm}\label{FIGURE1}
\end{minipage}%
}%
\subfigure{
\begin{minipage}[t]{0.35\linewidth}
\centering
\includegraphics[scale=0.35]{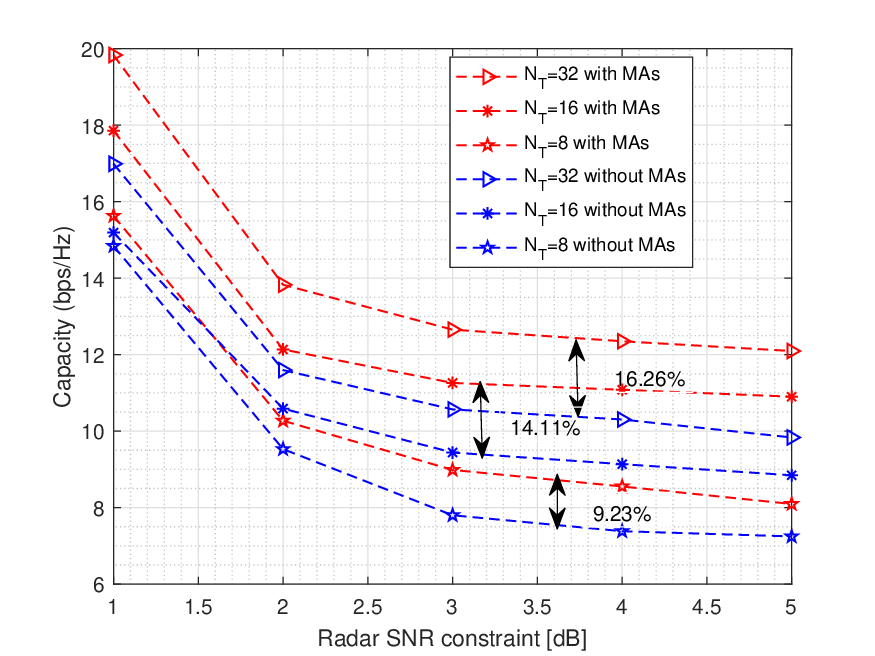}
\caption{Capacity versus sensing SNR}\label{FIGURE2}
\end{minipage}%
}%
\subfigure{
\begin{minipage}[t]{0.35\linewidth}
\centering
\includegraphics[scale=0.35]{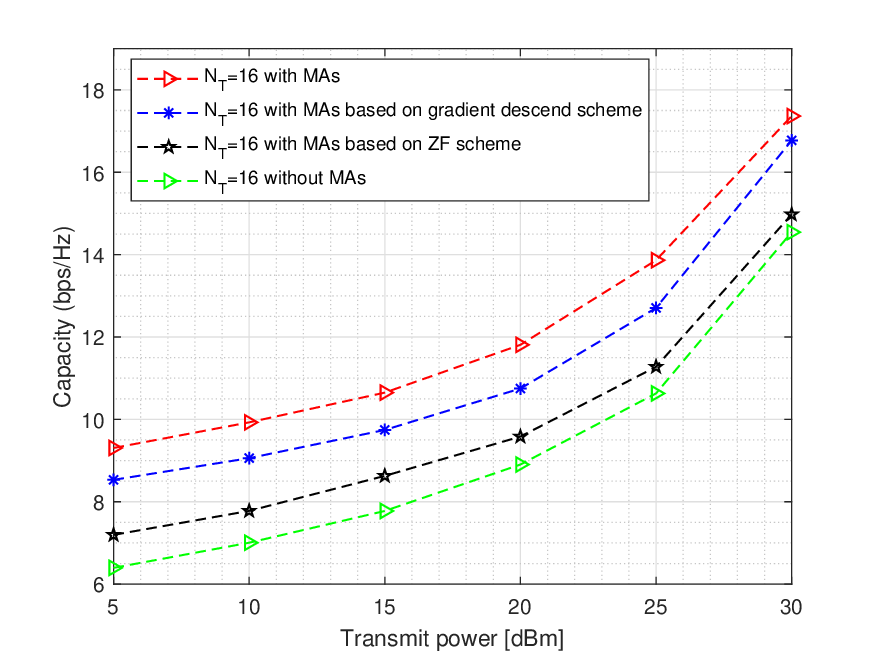}
\caption{Capacity versus transmit power}\label{FIGURE3}
\end{minipage}
}%
\centering
\end{figure*}

Fig.\ref{FIGURE1} illustrates the convergence pattern of the proposed algorithm. This shows that as the number of iterations increases, the system transmission rate increases steadily and converges in the 4-th to 8-th iterations. It is not difficult to find that as the number of MAs increases, the number of iterations for algorithm convergence increases. This is because with a larger number of MA positions to be optimized, the computation complexity will be increased. Furthermore, it can be observed that as the number of MAs increases, the transmission rate growth of MAs significantly outperforms that of systems without MAs. This is because when the number of antennas increases, the position freedom of MA increases.

Fig.\ref{FIGURE2} illustrates the relationship between transmission rate and radar sensing SNR. As the demand for sensing rate increases, the communication transmission rate decreases. This is because a larger proportion of the system's resource is allocated to meet the sensing SNR, causing the communication rate to decrease. In addition, it can be seen from the figure the transmission rate of the ISAC-MA system gradually approaches that without MA as the number of antennas decreases. This is because when the number of airborne beings is small, the position freedom of the MAs is reduced, and more transmission power is used to improve the sensing SNR, resulting in less power being used for communication transmission.

Fig.\ref{FIGURE3} illustrates the relationship between transmission rate and transmit power. As the transmission power increases, the communication rate increases. Moreover, compared with a conventional gradient descent optimization scheme and ZF waveform optimization scheme, the proposed algorithm shows better performance in terms of transmission rate. This is because our proposed algorithm uses a fourth-order Taylor approximation with lower error than the second-order Taylor approximation in the gradient descent algorithm. Compared with the ZF scheme, it does not consider the impact of sensing rate constraints.

\section{Conclusion and Future Works}\label{VI}
In this paper, we propose a novel MA-enabled ISAC system designed to increase the communication rate by optimizing the antenna positions at both the transmitter and receiver. We define a non-convex optimization problem. Then, we use mathematical approximations to transform it into a convex problem and perform an SDR method to obtain the final vector of optimal beamforming weights. Finally, we use SCA to optimize the transmit and receive MA positions. Simulation results show that the performance of the ISAC-MA system can be significantly improved as the number of MAs increases, resulting in a more favorable communication rate.

\end{document}